\documentclass{mn2e}
\usepackage[dvips]{graphicx}
\usepackage{amssymb}
\usepackage{times}

\newcommand{\sax}{{\it Beppo\-SAX}}
\newcommand{\source}{4U~1820--303}
\newcommand{\msun}{{\rm M}_{\sun}}

\newcommand{\xte}{{\it RXTE}}

\title[Orbital modulation of X-rays from \source]
{Dependence of the orbital modulation of X-rays from \source\ on the accretion rate}

\author[A. A. Zdziarski et al.]
{Andrzej A. Zdziarski,$^1$\thanks{E-mail:
aaz@camk.edu.pl} Marek Gierli\'nski,$^{2,3}$\thanks{E-mail: Marek.Gierlinski@durham.ac.uk} Linqing Wen,$^4$\thanks{E-mail: lwen@aei.mpg.de} and Zuzanna Kostrzewa$^5$\\
$^1$Centrum Astronomiczne im.\ M. Kopernika, Bartycka 18, 00-716 Warszawa, Poland\\
$^2$Department of Physics, University of Durham, Durham DH1~3LE, UK\\
$^3$Obserwatorium Astronomiczne Uniwersytetu Jagiello\'nskiego, Orla 171, 30-244 Krak{\'o}w, Poland\\
$^4$Max-Planck-Institut f\"ur Gravitationsphysik, Albert-Einstein-Institut, Am M\"uhlenberg 1, D-14476 Golm, Germany\\
$^5$Obserwatorium Astronomiczne Uniwersytetu Warszawskiego, Al.\ Ujazdowskie 4, 00-478 Warszawa, Poland
}

\date{Accepted 2007 February 28. Received 2007 January 16}

\pagerange{\pageref{firstpage}--\pageref{lastpage}}
\pubyear{2007}

\begin{document}

\maketitle

\label{firstpage}

\begin{abstract}  
We report the discovery, using \xte\/ data, of a dependence of the X-ray orbital modulation depth on the X-ray spectral state in the ultracompact atoll binary \source. This state (measured by us by the position on the X-ray colour-colour diagram) is tightly coupled to the accretion rate, which, in turn, is coupled to the phase of the $\sim$170-d superorbital cycle of this source. The modulation depth is much stronger in the high-luminosity, so-called banana, state than in the low-luminosity, island, state. We find the X-ray modulation is independent of energy, which rules out bound-free X-ray absorption in an optically thin medium as the cause of the modulation. We also find a significant dependence of the offset phase of the orbital modulation on the spectral state, which favours the model in which the modulation is caused by scattering in hot gas around a bulge at the disc edge, which both size and the position vary with the accretion rate. Estimates of the source inclination appear to rule out a model in which the bulge itself occults a part of an accretion disc corona. We calculate the average flux of this source over the course of its superorbital variability (which has the period of $\sim$170 d), and find it to be fully compatible with the model of accretion due to the angular momentum loss via emission of gravitational radiation. Also, we compare the dates of all X-ray bursts observed from this source by \sax\/ and \xte\/ with the \xte\/ light curve, and find all of them to coincide with deep minima of the flux, confirming previous results based on smaller samples.
\end{abstract}

\section{INTRODUCTION}
\label{intro}

\source\ is an ultracompact X-ray binary consisting of a low-mass He white dwarf accreting via Roche-lobe overflow onto a neutron star (e.g., Rappaport et al.\ 1987). The mass transfer occurs due to the loss of the angular momentum of the binary via emission of gravitational radiation (e.g., Rappaport et al.\ 1987; Zdziarski, Wen \& Gierli{\'n}ski 2007, hereafter ZWG07). Its orbital period of $P\simeq 685$ s was discovered in X-rays as a modulation with a $\sim$2 per cent peak-to-peak amplitude (Stella, Priedhorsky \& White 1987). The period is very stable, with a low $\dot P/P=(-3.5\pm 1.5)\times 10^{-8}$ yr$^{-1}$ (Chou \& Grindlay 2001, hereafter CG01), which makes it certain it is due to the binary motion. The $\dot P<0$ appears to be in conflict with the gravitation-radiation induced mass transfer from the light white dwarf to the heavier neutron star (Rappaport et al.\ 1987), but it may possibly be due to gravitational acceleration of the system in the parent globular cluster NGC 6624 (CG01). 

The source distance has been estimated as $D=6.4\pm 0.6$ kpc, 7.4--7.6 kpc, $8.6\pm 0.4$ kpc, and $7.6\pm 0.4$ kpc based on calibration of horizontal-branch stars in NGC 6624 by Vacca, Lewin \& van Paradijs (1986), Rich, Minniti \& Liebert (1993), Richtler, Grebel \& Seggewiss (1994), and Kuulkers et al.\ (2003; based on the {\it HST\/} results of Heasley et al.\ 2000), respectively. Kuulkers et al.\ (2003) found $D=7.6\pm 0.4$ kpc compatible with the peak X-ray burst luminosity being a standard candle in their sample of sources, and equal to the Eddington limit for He. Then, Cumming (2003) found $D=7.6$ kpc consistent with his theoretical modelling of X-ray bursts in \source. On the other hand, Shaposhnikov \& Titarchuk (2004) claimed $D=5.8$ kpc based on their model of a single X-ray burst observed from \source\/ by {\it Rossi X-ray Timing Explorer\/} (\xte). 

A very unusual feature of \source\ is the {\it intrinsic\/} luminosity variation by factor of $\ga$2 at a (superorbital) period of $P_{\rm s}\simeq 170$ d (Priedhorsky \& Terrell 1984; Smale \& Lochner 1992; CG01; {\v S}imon 2003; Wen et al.\ 2006; ZWG07). CG01 found the modulation to be stable with $P_{\rm s}=171.0\pm 0.3$ d and $\vert \dot P_{\rm s}/P_{\rm s}\vert <2.2\times 10^{-4}$ yr$^{-1}$ based on data from 1969 to 2000. This value of $P_{\rm s}$ is also compatible with the \xte\/ All Sky Monitor (ASM) data from 1996 through 2006 (Wen et al.\ 2006; ZWG07). X-ray bursts from the source take place {\it only\/} around the flux minima (Cornelisse et al.\ 2003; Section \ref{data} below), which proves that the observed variability is indeed due to intrinsic accretion rate changes.  This is further supported by strong correlations between the observed flux and the source spectral state, varying with the flux in a way typical of atoll-type neutron-star binaries (Bloser et al.\ 2000; Gladstone, Done \& Gierli{\'n}ski 2007), and between the frequency of kHz QPOs observed from the source and the flux (Zhang et al.\ 1998; van der Klis 2000). 

To explain the long-term periodic variability of the accretion rate, CG01 proposed a hierarchical triple model, in which a distant tertiary exerts tidal forces on the inner binary. This so-called Kozai process results in quasi-periodic changes of the eccentricity of the inner system (e.g., Kozai 1962; Mazeh \& Shaham 1979; Ford, Kozinsky \& Rasio 2000; Wen 2003). This causes changes of the distance between the inner Lagrange point, $L_1$, and the center of mass of the donor, changing, in turn, the accretion rate. This model has been elaborated in detail by ZWG07 and found fully compatible with the observational data. 

Here, we study the possibility of a dependence of the depth of the X-ray modulation on the accretion rate (related to the superorbital phase). There are a number of reasons to predict such dependence. First, the modulation may be due to obscuration/scattering of the X-rays by a structure at the disc edge forming around the point at which the accretion flow from the companion impacts the outer edge of the disc, as proposed by Stella et al.\ (1987). Similar obscuration is present  in a number of other low-mass X-ray binaries, e.g., in another short-period, white-dwarf--neutron-star system, 4U 1916--053 (White \& Swank 1982). Then, the parameters of the edge structure will, most likely, depend on the accretion rate. These size changes may, in turn, change the depth of the modulation. Furthermore, the flow through the $L_1$ point in an eccentric binary is modulated at the orbital period. The modulation amplitude depends on the value of the eccentricity, which, as discussed above, is most likely variable in \source. Though smoothed during the viscous flow through the disc, a residual modulation may still occur in the X-ray producing region near the surface of the neutron star. Finally, the inner binary plane precesses during the Kozai process, which effect may be observable if the mutual inclination of the outer and inner binaries is low (ZWG07). If the orbital modulation is due to obscuration/scattering, precession of the binary plane is very likely to affect its amplitude. 

We test and confirm the above prediction using the \xte\/ Proportional Counter Array (PCA) and High Energy X-Ray Timing Experiment (HEXTE) data. We also estimate the average bolometric flux from this source, and present a comparison of the dates of all X-ray bursts observed by \sax\/ and \xte\/ with the \xte\/ light curve. Also, we discuss contraints on the source inclination.

\section{The data}
\label{data}

\begin{figure*}
\centerline{\includegraphics[width=145mm]{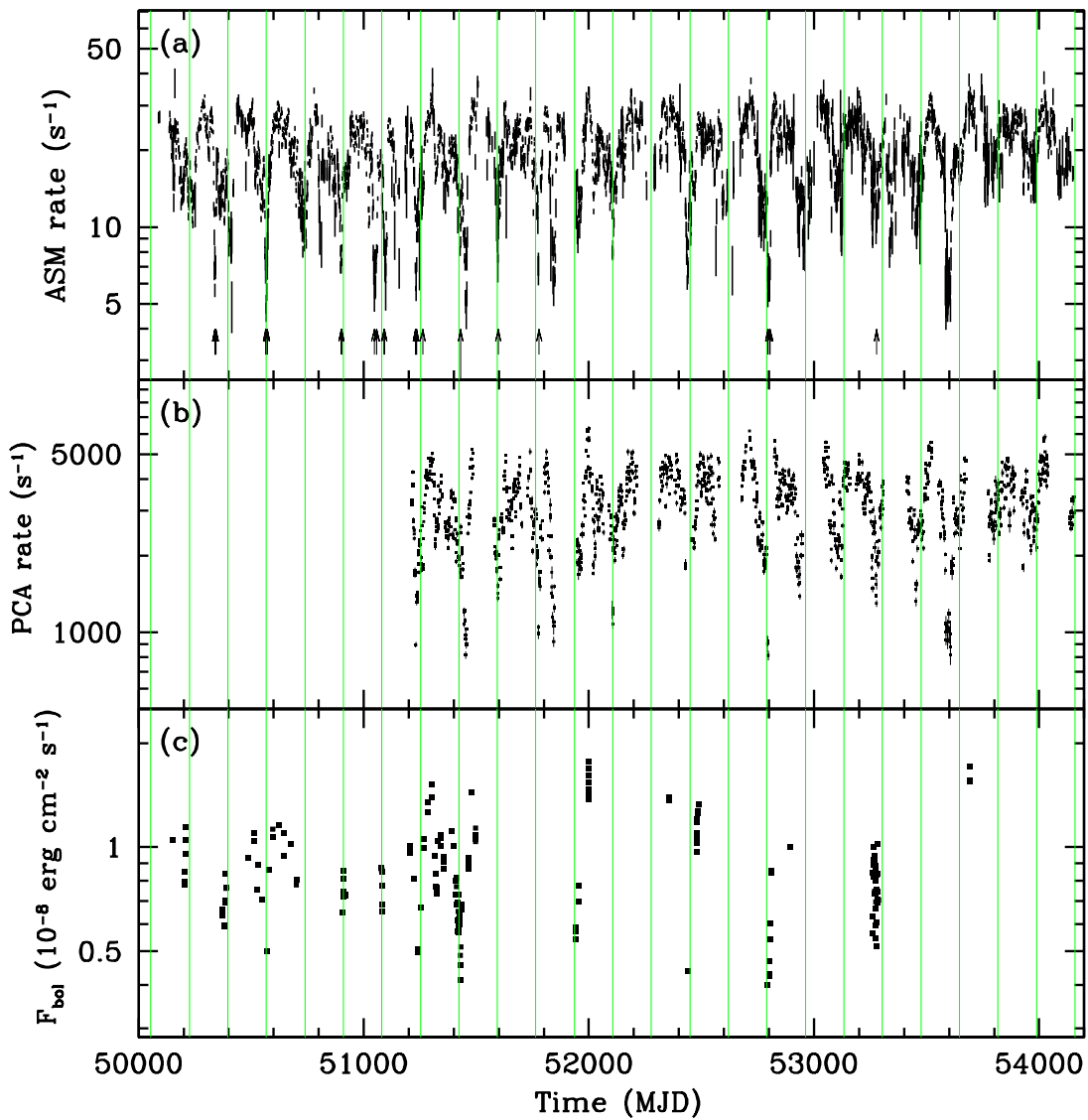}}
\caption{(a) The \xte/ASM light curve based on 1-day average measurements, rebinned to the minimum significance of $3\sigma$. The short arrows shows the times of X-ray bursts, and the long arrow indicates the X-ray burst followed by the superburst. (b) The PCA light curve based on Galactic bulge scans. (c) The light curve of the bolometric flux calculated from the PCA and HEXTE data. The vertical lines show the minima (the phase $\phi_{\rm s}=0$) of the superorbital cycle according to the ephemeris of CG01. See Section \ref{data}.
\label{lc} }
\end{figure*}

Fig.\ \ref{lc}(a) shows the \xte\/ ASM light curve of \source\ for 1996 January 1--2007 February 20 (the same as used in ZWG07). Fig.\ \ref{lc}(b) shows the \xte\/ PCA data from Galactic bulge scans\footnote{lheawww.gsfc.nasa.gov/users/craigm/galscan/html/4U\_1820-30.html}(also used in ZWG07) for the time interval of 1999 February 5--2007 February 24. The vertical lines show the minima of the superorbital cycle according to the ephemeris of CG01. The arrows in Fig.\ \ref{lc}(a) show the times of X-ray bursts from this source observed by the \sax\/ Wide Field Camera (Cornelisse et al.\ 2003)\footnote{Those bursts occured on MJD 50339 (1), 50340 (2), 50341 (2), 50342 (2), 50343 (1), 50564 (5), 50572 (2), 50900 (2), 50901 (1), 51047 (3), 51057 (4), 51058 (1), 51091 (3), 51092 (4), 51230 (3), 51231 (4), 51232 (2), 51233 (3), 51234 (1), 51262 (1), 51597 (1), where the number in parentheses gives the number of bursts detected in a given day (Jean in 't Zand, private communication).} and the \xte\/ PCA (Galloway et al.\ 2007). Beside the 4 PCA bursts listed by Galloway et al.\ (2007), we also show the X-ray burst that triggered the 3-hr superburst observed from this source (MJD 51430, Strohmayer \& Brown 2002), and an additional one found during our analysis of the PCA data (see below), on MJD 53277\footnote{In total, the PCA detected X-ray bursts on MJD 50570, 51430, 52794, 52802, 52805, 53277.}. We see that {\it all\/} of the X-ray bursts took place in deep minima of the X-ray flux (as pointed out by Cornelisse et al.\ 2003 for the \sax\/ data), confirming the intrinsic character of the observed variability. 

We then study the \xte\/ spectral data from the PCA (detector 2, top layer only) and HEXTE (both clusters) detectors, extracted using {\sc ftools} v.\ 5.3, and with time intervals containing the X-ray bursts removed. We use all publically-available observations between 1996 March 9 and 2005 November 19. We use a spectral model (normalized to the PCA data) similar to that of Done \& Gierli{\'n}ski (2003), consisting of a multicolour disc blackbody (Mitsuda et al.\ 1984) and thermal Comptonization (Zdziarski, Johnson \& Magdziarz 1996) absorbed by a medium with $N_{\rm H}=2.4\times 10^{21}$ cm$^{-2}$ (Bloser et al.\ 2000). In addition, a smeared edge (approximating effects of reflection) and a Gaussian line were included. However, unlike Done \& Gierli{\'n}ski (2003), we allow the temperature of the seed photons for Comptonization to be different from the maximum temperature of the blackbody disc, which we have found to be required by the soft-state data. Fig.\ \ref{lc}(c) shows the resulting light curve of the intrinsic bolometric flux from the fitted model, $F_{\rm bol}$. 

\begin{figure}
\centerline{\includegraphics[width=84mm]{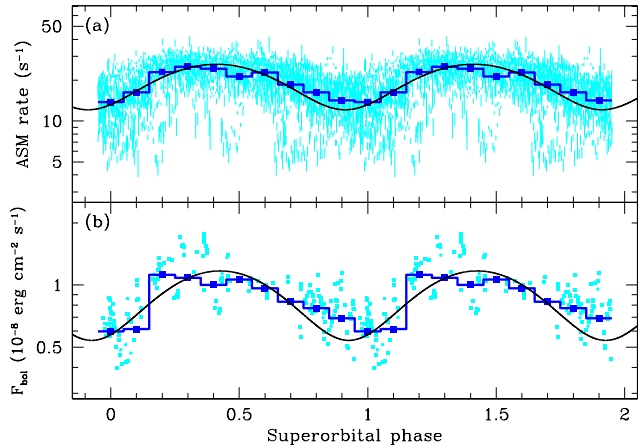}}
\caption{(a) The ASM light curve of Fig.\ \ref{lc}(a) folded on the superorbital ephemeris of CG01. (b) The folded light curve for the bolometric flux (see Section \ref{data}). On each panel, the histogram shows the light curve averaged over 10 phase bins, and the solid curve shows the theoretical model of ZWG07 fitted to the averaged data. 
\label{phase} }
\end{figure}

Fig.\ \ref{phase}(a) shows the light curve of the ASM count rate folded on the superorbital ephemeris of CG01. The histogram shows then the light curve averaged within 10 phase bins, and the solid curve shows the theoretical model of ZWG07 fitted by them to these data. Fig.\ \ref{phase}(b) shows the same for $F_{\rm bol}$. Here, we use the same model as ZWG07, and fit the normalization and the phase offset to the $F_{\rm bol}$ phase-folded and averaged within each bin. We find the phase offset to be $\phi_{\rm s,0}/2\upi =-0.074$, very similar to that for the PCA scanning data (ZWG07).

Based on the data and the spectral model, the average flux is $\langle F_{\rm bol} \rangle \simeq (8.7\pm 0.2)\times 10^{-9}$ erg cm$^{-2}$ s$^{-1}$, which corresponds to the average isotropic luminosity of $\langle L\rangle \simeq (6.0\pm 0.2)\times 10^{37} (D/7.6\,{\rm kpc})^2$ erg s$^{-1}$. The stated uncertainty of the average is only statistical, equal to the standard deviation of $F_{\rm bol}$ divided by the square root of the number of measurements (and it does not take into account a systematic uncertainty related to the flux calibration of the PCA and that related to the adopted spectral model). As discussed in ZWG07, this luminosity is fully compatible with the standard theoretical mass-transfer rate of Rappaport et al.\ (1987). Similarly to the result of ZWG07 based on the ASM and scanning PCA data, we find the dynamic range of the folded/averaged $F_{\rm bol}$ to be by a factor of $\sim$2.

\begin{figure}
\centerline{\includegraphics[width=84mm]{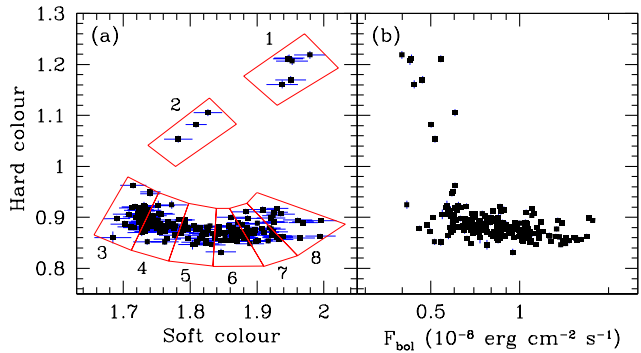}}
\caption{(a) The intrinsic colour-colour diagram, as given by the ratio of the energy fluxes in the photon energy ranges of (9.7--16 keV)/(6.4--9.7 keV) and (4.0--6.4 keV)/(3.0--4.0 keV) for the hard colour and soft colour, respectively. 
The boxes show our division into 8 spectral substates. (b) The corresponding dependence of the hard colour on the bolometric flux.
\label{col_col} }
\end{figure}

We then use the fit results to calculate the intrinsic colour-colour diagram, i.e., based on physical, absorption-corrected, fluxes in four energy bands, shown in Fig.\ \ref{col_col}(a). We divide the data into 8 spectral substates, with the substates 1--2 and 3--8 corresponding to the so-called (Hasinger \& van der Klis 1989) island and banana states, respectively. Fig.\ \ref{col_col}(b) shows the dependence of the hard colour on $F_{\rm bol}$. Although we see here some overlap in $F_{\rm bol}$ between the island and banana states, $F_{\rm bol}$ generally increases with the substate number. Also, Bloser et al.\ (2000) found that when {\it individual\/} superorbital cycles are considered, the position on the colour-colour diagram follows the superorbital phase. 

For each pointed observation, we have also extracted its light curve from the PCA Standard-2 data, detectors 0 and 2, all layers, in 16-s bins, in the full 2--60 keV PCA energy band. We have subtracted background from the light curves and applied barycentric corrections.

\section{Variability of the orbital modulation}
\label{modulation}

We can see in Fig.\ \ref{lc}(c) that the coverage of the \source\ light curve by pointed \xte\/ observations is relatively sparse, as well the flux varies within a factor of up to $\sim$3 for a given superorbital phase, $\phi_{\rm s}$. This agrees with the \xte\/ ASM light curve (Fig.\ \ref{lc}a), where we see that although the superorbital ephemeris of CG01 gives a generally good prediction of the flux minima, in several cases the minima occur tens of days away from the prediction, as well as there are a number of additional minima, indicating the presence of strong aperiodic variability in addition to the periodic one (cf.\ {\v S}imon 2003). A more detailed view of the ASM light curve is given in fig.\ 1 of ZWG07. Thus, the value of $\phi_{\rm s}$ is {\it not\/} an accurate enough criterion to determine the actual stage of the superorbital cycle corresponding to a given pointed observation. Still, we have divided the available pointed observations into 8 bins based on $\phi_{\rm s}$ (Figs.\ \ref{lc}c, \ref{phase}b), in order to find the average amplitude of the orbital modulation as a function of that phase. However, the distribution of the PCA count rate, $R$, within a given phase bin is highly non-Gaussian, with, e.g., three individual peaks in the bin centered on $\phi_{\rm s}=0$. We have also found that those data correspond to a wide range of spectral colours. 

Therefore, for the purpose of studying average orbital modulation, we remove the variability on time scales much longer than the orbital period, $P$. This is done by fitting a straight line to each uninterrupted part of the light curve contributing to a given superorbital bin and then dividing the light curve by that dependence. However, we do not take into account those resulting pieces of the light curve that have the length $<P$. Within each superorbital bin, we fold and average the renormalized PCA light curves (with the detrended count rate, $R_{\rm d}$) on the orbital period using the quadratic ephemeris given by eq.\ (4) of CG01. We then fit the the detrended count rate, $R_{\rm d}$, by a sinusoidal dependence, $A \cos (\phi - \phi_0) + C$, where $\phi$ is the orbital phase given by the ephemeris, and the constant component, $C$ ($\simeq 1$ after the renormalization), the modulation amplitude, $A$, and the offset phase, $\phi_0$ are free parameters. This method is used hereafter.

We have then found only a weak dependence of the relative modulation depth, $A/C$, on the superorbital phase, with the value for the bin centered on $\phi_{\rm s}=0$ of 0.008 compared to $\simeq 0.01$ in other bins. As discussed above, this result does not rule out an actual strong depedence of $A/C$ on the phase in an {\it single\/} superorbital cycle because of the relatively large scatter of the duration of each cycle and the presence of additional minima and strong aperiodic veriability in general. Those effects result in a large scatter of the fluxes (and the spectral states) contributing to a given phase bin. Also, the statistics of the PCA data contributing to a given superorbital bin is rather limited. We have found, in addition, $\phi_0/2 \upi\simeq -0.07\simeq$ constant. The fact that $\phi_0<0$ in the average PCA data indicates the long-term decrease of $P$ being slightly faster than that predicted by the ephemeris of CG01. We also note that CG01 used heliocentric time with some approximations, while we strictly use the barycentric time.

We have then used the PCA count rate as the criterion, dividing the data into 8 bins, with the average rate before the renormalization, $\langle R\rangle$ within each bin spanning the range from about 500 s$^{-1}$ to 2700 s$^{-1}$. In that case, the fractional modulation amplitude increases from $A/C\simeq (5$--$8)\times 10^{-3}$ in the first two count-rate bins to $\simeq 0.01$ at higher $\langle R\rangle$, and increasing to the rather high value of $\simeq (2.1\pm 0.2)\times 10^{-2}$ at the highest $\langle R\rangle$. 

\begin{figure}
\centerline{\includegraphics[width=84mm]{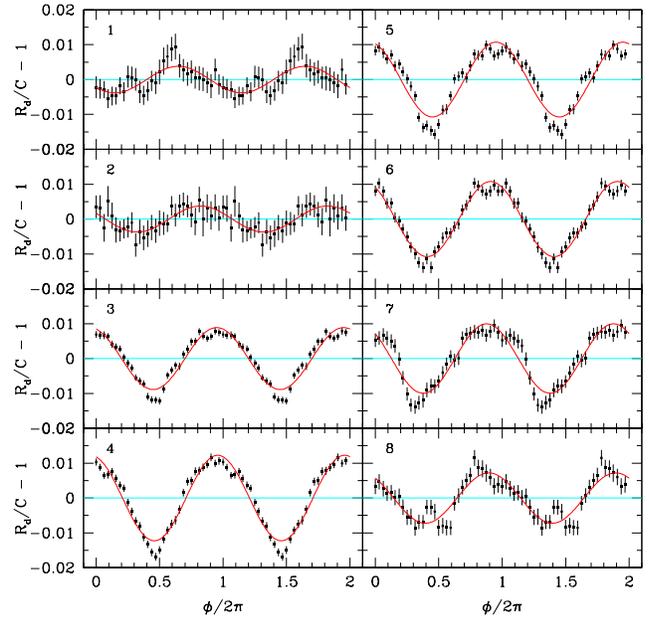}}
\caption{The fractional orbital modulation shown for the 8 spectral substates defined in Fig.\ \ref{col_col}. The error bars on $R_{\rm d}/C$ represent the standard deviation within the data set contributing to a given phase bin.
\label{mod_state} }
\end{figure}

Then, we have studied the dependence of the fractional modulation on the spectral state, which results we show in detail. Fig.\ \ref{mod_state} shows the fractional orbital modulation profiles, $R_{\rm d}/C-1$ (i.e., after removing the long-term variability, see above), for the 8 spectral substates defined in Fig.\ \ref{col_col}. We see a clear difference between the low orbital modulation in the island state (substates 1--2) and the banana state (substates 3--8). The island state, corresponding to the lowest accretion rates, has a much weaker fractional orbital modulation. We also note that the modulation is more complex than sinusoidal in some cases. 

\begin{figure}
\centerline{\includegraphics[width=84mm]{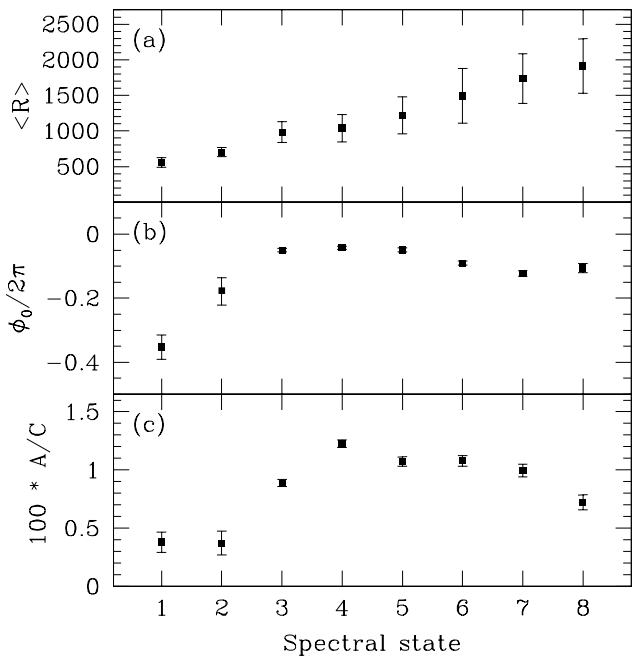}}
\caption{ The average PCA count rate, the offset phase, and the amplitude of the fractional modulation vs.\ the spectral substate (defined in Fig.\ \ref{col_col}). The error bars on $\langle R\rangle$ represent the standard deviation, not the error of the average (which are negligibly small). The errors on $\phi_0/2\upi$ and $A/C$ correspond to 90 per cent confidence intervals for the fits. 
\label{fit_state} }
\end{figure}

Fig.\ \ref{fit_state} quantifies our results. We see that $\langle R\rangle$ increases with the spectral state number, consistent with the increasing accretion rate. Then, the differences in $A/C$ between the island and banana states are much larger than the errors. Therefore, we establish the dependence of the depth of the orbital modulation on the spectral state, and thus on the accretion rate, at very high confidence. We also note that though the fractional modulation first strongly increases from the island into the banana state, it then somewhat decreases at the extreme banana state, indicating a further complexity of this dependence. We also see a marked increase of the offset phase from the island state to the banana state.

\begin{figure}
\centerline{\includegraphics[width=70mm]{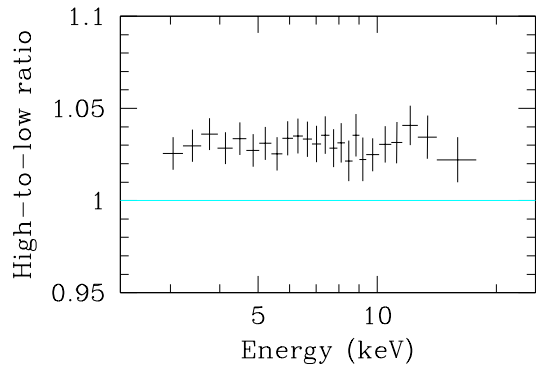}}
\caption{ The ratio between the spectra corresponding to the maxima and minima of the orbital modulation for the spectral substate 4. See Section \ref{modulation} for details.
\label{ratio} }
\end{figure}

Then, we have looked into the dependence of the orbital modulation on energy. To study it, we have obtained spectra corresponding to the maximum and minimum of the modulation. We used the spectral state 4, where the amplitude of the modulation is highest, and formed spectra for the 4 points around each the maximum and the minimum (Fig.\ \ref{mod_state}). Fig.\ \ref{ratio} shows the ratio between the two spectra. We see it is constant to a rather high accuracy. This shows the modulation cannot be due to bound-free X-ray absorption, as it strongly decreases with energy. The constant ratio can be either due to scattering by an ionized medium with the bound-free absorption negligible at $\ga$3 keV (e.g., an accretion disc corona) or partial obscuration by a Thomson-thick medium.

\section{Interpretation}
\label{interpretation}

\subsection{Inclination}
\label{inclination}

The value of the inclination, $i$, between the normal to the binary plane and the line of sight is of major importance for constraining models of the observed orbital modulation. We first derive the upper limit, $i_{\rm max}$, due to the lack of eclipses. The radius of the white dwarf, $R_2$, can be derived by combining the mass-radius relation for He white dwarfs, the Kepler law and the Roche-lobe radius formula (Rappaport et al.\ 1987; see also ZWG07). Here, we consider the range of mass of the neutron star of $M_1=(1.28$--$1.5)\msun$, and of the white dwarf mass of $M_2= (0.057$--$0.077) \msun$, which corresponds to its radius being between 0 and 20 per cent higher than that of the fully degenerate He configuration (Zapolsky \& Salpeter 1969), which corresponds in turn to $R_2= (2.08$--$2.29)\times 10^9$ cm. For the above ranges of masses, the separation (from the Kepler law) is $a=(1.28$--$1.36)\times 10^{10}$ cm. Then, $R_2/a\simeq (0.15$--$0.18)$ and $i_{\rm max} =90\degr - \arcsin(R_2/a) \simeq (80$--$81)\degr$. 

Anderson et al.\ (1997) have derived $i=43^{+8}_{-9}\degr$ by interpreting the orbital modulation discovered by them in the UV as due to irradiation of the white dwarf by the X-ray source, using the theoretical model of Arons \& King (1993). However, their stated uncertainty is only statistical, and neglects systematic uncertainties of the model, in particular, they neglect shadowing of the white dwarf by the accretion disc. Including that effect would move the allowed range of $i$ to higher values than that above. 

On the other hand, Compton reflection (e.g., Magdziarz \& Zdziarski 1995) and a fluorescent Fe K$\alpha$ emission line were discovered in the X-ray spectra of the superburst from this source (Strohmayer \& Brown 2002). It was then theoretically interpreted as due to irradiation by the X-rays from the neutron star surface of the surrounding accretion disc by Ballantyne \& Strohmayer (2004). These spectral features were relatively strong, e.g., the line equivalent width of $\sim$70--200 eV (Strohmayer \& Brown 2002), which, provided the reflection is indeed from a flat disc in the binary plane, rules out inclinations close to edge on. Ballantyne \& Strohmayer (2004) have also fitted the reflection inclination using the relativistic broadening of the reflection features. They obtained a compatible result of $i<36\degr$ at the $2\sigma$ significance, which is marginally compatible with the constraint of $i>34\degr$ of Anderson et al.\ (1997). Both those results strongly favour intermediate inclinations, ruling out $i$ close to the eclipse limit.

\subsection{Models}
\label{models}

Stella et al.\ (1987) proposed that the X-ray modulation is due to occultation by a structure at the edge of the accretion disc, noting that such structure is thickest at the point where the gas stream impacts. In the first of their two models, the structure is optically thin and strongly ionized. Then, scattering by that medium would remove some photons from the line of sight. In the second model, an accretion disc corona above the disc (but not the primary X-ray source itself) is partially obscured by an optically-thick disc rim (which is highest at the point of impact of the gas stream), analogously to models of dipping low-mass X-ray binaries (e.g., White \& Swank 1982; Hellier \& Mason 1989). In both cases, the height and form of the edge structures is highly likely to depend on the accretion rate, explaining the dependence found by us (Section \ref{modulation}).

In both models, the height/radius ratio of the occulting structure has to be more than $R_2/a$, which is constrained to be in the range of 0.15--0.18 (Section \ref{inclination}). This appears to be possible, e.g., the largest height/radius ratio of an optically-thick disc rim found by Hellier \& Mason (1989) is $\simeq$0.22. Still, this value corresponds to $i\ga 76\degr$, which appears ruled out for \source\ (Section \ref{inclination}).

On the other hand, the disc rim may be surrounded by an optically-thin, highly ionized, gas, with a higher scale height. This gas may, in fact, connect to the corona above the outer parts of the disc. The difference in the Thomson optical depth of $\sim$0.02 in the lines of sights corresponding to the orbital phases of $\phi/2\upi= 0$ and 0.5 can explain the modulation. Then, both the optical depth and the nodal angle of that gas are likely to depend on the accretion rate. The latter can then explain the dependence of the offset phase of the orbital modulation on the spectral state. 

Also (as discussed in ZWG07), the binary plane may measurably precess due to the influence of the tertiary, depending on the relative inclination between the inner and outer binary planes (which remains not determined). This would change the binary inclination, $i$, which is also likely to affect the strength of any absorption/scattering process. However, this process alone is not expected to yield any changes of the offset phase, contrary to our results. Still, a binary plane precession may take place in addition to changes of the occulting structure.

In addition, the presence of eccentricity will cause a modulation of the flow rate through the $L_1$ point over the course of an orbit. As calculated by ZWG07, this flow rate will be variable by a factor of $\sim$10--20 near the maximum of the eccentricity, i.e., around the maxima of the superorbital cycle (and much less at the minima of the cycle). This modulation will, however, be damped over the viscous time, $t_{\rm vis}$, it takes for the accreted matter to flow from its outer rim to the inner one. In the standard cold accretion model (Shakura \& Sunyaev 1973), $t_{\rm vis} \sim 10^5$ s for the parameters of \source, in which time any plausible dispersion of the inflow speed will probably make the modulation of the $L_1$-flow unobservable in the X-rays, which are emitted close to the neutron star.  

However, Gilfanov \& Arefiev (2007) found breaks in the power spectra in a number of X-ray binaries at frequencies proportional to the orbital frequencies. They interpreted those frequencies as occuring at $t_{\rm vis}^{-1}$, which then implied that $t_{\rm vis}$ were much shorter than in the standard model, which they explained by the presence of hot coronal flows. In particular, they claimed $t_{\rm vis}\sim 300\, {\rm s}\sim P/2$ in \source. After such a short time interval, most of the flow-rate modulation at the $L_1$ would still be present close to the neutron star surface. Then most of the large modulation amplitude, $\ga 10$, present around the maxima of the superorbital cycle (see above) should be visible, but it is clearly not seen. Thus, if the superorbital modulation is due to the variable eccentricity (as successfully modelled by ZWG07), $t_{\rm vis}\sim 300$ s is ruled out for \source. [We note that Gilfanov \& Arefiev (2007) found the break in the power spectrum of \source\ to look significantly different from those in other X-ray binaries of their sample, and thus our result does not rule out their general interpretation of the breaks as equal to $t_{\rm vis}^{-1}$.] Then, there could be some intermediate value of $t_{\rm vis}$ which would explain the observed strength orbital modulation. However, it is again not clear how to explain the observed variable offset phase in this model. 

We note that while a modulation due to obscuration or scattering depends on the binary phase (e.g., with respect to the superiour conjunction of the secondary), a modulation due to the eccentricity would depend on the phase with respect to the periastron, which position precesses within a superorbital cycle. In the former case, the measured period of the X-ray modulation is exactly the orbital period, whereas for the latter case, it is a beat between the orbital rotation and the periastron precession. However, given that $P_{\rm s}/P\simeq 2.2\times 10^4$, this effect appears difficult to measure. 

\section{Comparison with other sources}
\label{comparison}

As discussed by ZWG07, the intrinsic, accretion-rate related, character of the long-term periodic flux changes in \source\ appears unique among all sources in which superorbital variability has been discovered so far. In other cases, the changes of the observed flux are thought to be caused by accretion disc and/or jet precession (e.g., in Her X-1, SS 433, Cyg X-1; Katz 1973, 1980; Lachowicz et al.\ 2006). However, in that precession the inclination angle of the disc with respect to the orbital plane is constant, and thus it cannot change the accretion rate (or the luminosity). Also, the ratio between the superorbital and orbital periods is $\simeq 2.2\times 10^4$, which is much higher than that possible to obtain from any kind of disc precession at the mass ratio of the system (e.g., Larwood 1998; Wijers \& Pringle 1999). 

On the other hand, a dependence of the profile of the X-ray orbital modulation on the spectral state and/or on the superorbital phase has been found in two other X-ray binaries showing  both orbital and superorbital modulations. Scott \& Leahy (1999) and Ibragimov, Zdziarski \& Poutanen (in preparation) found the profiles of the orbital modulation to depend on the superorbital phase in Her X-1 and Cyg X-1, respectively. In both cases, the interpretation of these effects is geometrical, via the dependence of the observed flux on the viewing angle of a precessing disc and associated changes of the obscuring medium. In Cyg X-1, this effect leads to the minimum of orbital modulation (caused by bound-free absorption in the wind from the companion, Wen et al.\ 1999) at the maximum of the superorbital cycle, i.e., when the disc is viewed most face-on and the absorption is weakest. This is opposite to the dependence found in \source, where it is caused by the varying accretion rate. 

We have also looked in the \xte/ASM data for a dependence of the orbital modulation on the superorbital phase in other binaries showing both periodicities, namely 2S 0114+650, LMC X-4, and SS 433 (see Wen et al.\ 2006). Unfortunately, the statistics of the data in all cases was insufficient to either confirm or reject the presence of this effect. However, we have not looked at the dependence of the orbital modulation directly on X-ray flux or X-ray hardness (found to be dominant in \source, see Section \ref{modulation}). For Her X-1, we have confirmed the presence of this dependence and its stationarity in the 1996--2006, $\sim$3,800-d interval, ASM data (as compared to the $\sim$800-d interval analyzed by Scott \& Leahy 1999). 

\section{Conclusions}

We have analyzed the currently available \xte\/ PCA/HEXTE spectral data. From broad-band spectral fitting with a physically-motivated model, we have determined the average bolometric flux as $\langle F_{\rm bol}\rangle\simeq 8.7\times 10^{-9}$ erg cm$^{-2}$ s$^{-1}$. At the most likely distance of $D\simeq 7.6$ kpc, the corresponding bolometric luminosity fully agrees with the detailed model of R87, in which the accretion proceeds via a Roche-lobe overflow due to the angular momentum loss via emission of gravitational radiation. 

We have found that the strength of X-ray orbital modulation in \source\ depends on the accretion rate, as measured by either the spectral state (as given by the position on the colour-colour diagram) or the X-ray flux, and which is closely related to the superorbital phase of this source. The relative strength of the modulation is about twice as high in the high-accretion rate, banana state than in the low-accretion rate, island state. 

We also see a dependence of the phase of the modulation on the spectral state. The phase dependence favours the explanation of the dependence as due to scattering by a hot gas around the accretion-stream bulge at the outer disc edge, which both the size and position are likely to depend on the accretion rate. Alternative explanations are precession of the binary plane during the superorbital Kozai cycle, and the residual modulation of the rate of accretion onto the neutron star due to the (superorbital-phase dependent) eccentricity of the orbit. However, it is not clear how to explain the dependence of the phase of the orbital modulation on the spectral state in those models. Also, occultation of an accretion disc corona by the bulge itself appears to be ruled out by estimates of the source inclination.

We also find that the orbital modulation is approximately independent of energy. This rules out the model of the modulation as due to bound-free absorption in an accetion disc corona, and it shows that the process causing the modulation is itself energy-independent.

\section*{ACKNOWLEDGMENTS}

We thank J. in 't Zand for providing us with the dates of the X-ray bursts observed by \sax\/ from \source, and R. Misra for valuable discussions. This research has been supported in part by the Polish grants 1P03D01827, 1P03D01128 and 4T12E04727. MG acknowledges support through a PPARC PDRF. LW acknowledges support by the Alexander von Humboldt Foundation's Sofja Kovalevskaja Programme (funded by the German Federal Ministry of Education and Research). The work of ZK was done at the Copernicus Center during the 2006-summer undergraduate student program. We also acknowledge the use of data obtained through the HEASARC online service provided by NASA/GSFC.

\label{lastpage}
\end{document}